\documentclass[aps,prl,twocolumn,showpacs,amsmath,amssymb,10pt,notitlepage]{revtex4-1}%
\usepackage{graphicx}
\usepackage{dcolumn}
\usepackage{epsfig,bm}
\usepackage{amsfonts}
\usepackage{color}
\usepackage{bigints}

\begin{document}

\title{
Non-mean-field effects in systems with long-range forces in competition
}
\author{R. Bachelard}
\affiliation{Instituto de F\'isica de S\~ao Carlos, Universidade de S\~ao Paulo, 13560-970 S\~ao Carlos, SP, Brazil}
\affiliation{University of Nova Gorica, School of Applied Sciences, Vipavska 11c SI-5270 Ajdov\v{s}\v{c}ina, Slovenia}

\author{F. Staniscia}
\affiliation{Max Planck Institute for the Physics of Complex Systems, 01187 Dresden, Germany}
\affiliation{University of Nova Gorica, School of Applied Sciences, Vipavska 11c SI-5270 Ajdov\v{s}\v{c}ina, Slovenia}

\date{\today}

\begin{abstract}
We investigate the canonical equilibrium of systems with long-range forces in competition. These forces create a modulation in the interaction potential and modulated phases appear at the system scale. The structure of these phases differentiate this system from monotonic potentials, where only the mean-field and disordered phases exist. With increasing temperature, the system switches from one ordered phase to another through a first-order phase transition. 
Both mean-field and modulated phases may be stable, even at zero temperature, and the long-range nature of the interaction will lead to metastability characterized by extremely long time scales.
\end{abstract}

\maketitle

Long-range interactions are present in a wide variety of physical systems such as gravitation, plasma physics, wave-particle interactions or two-dimensional fluids~\cite{Campa09}. The dynamics and thermodynamics of these systems is dominated by a mean-field phase where all elements in the system are in a similar phase (e.g. polarization of spins). In lattices with long-range forces, this result was confirmed in microcanonical simulations~\cite{Anteneodo98,Tamarit00}, and rigorously for microcanonical and canonical ensembles~\cite{Campa00,Campa03}. These results validate the mean-field approach for long-range systems.

The scenario changes when a short-range force enters into competition with the long-range one. It was shown that stripped patterns and other modulated phases may emerge~\cite{Nagle70,Low94,Cannas04,Mukamel05,Giuliani06,Giuliani07,Dauxois10,Edlund10}. A similar scenario occurs in the case of competing short-range forces: superconducting films, magnetic systems, chemical reactions (i.e. Turing patterns) and copolymers~\cite{Seul95}. Short-range purely repulsive forces can also generate such a phenomenon~\cite{Malescio03}.

Systems with two competing long-range forces are investigated here, producing a situation where phases modulated at the size of the system appear. This study is motivated by proposals of tabletop experiments where long-range forces are created artifically using  cold atoms or colloids~\cite{ODell00,Dominguez10,Golestinian12}. For example, Chalony et al. recently took advantage of laser-induced atom-atom interaction to create an artificial one-dimensional gravity~\cite{Chalony12}. These works open new perspectives to study, and tune, long-range forces. Modulated long-range potentials are also naturally found in the interaction betweem light and ultracold matter~\cite{Courteille10,Bachelard12}, making these systems appealing for future work.

In this paper, it is shown that the competition of long-range forces in a lattice leads to canonical equilibria with phases modulated at the scale of the system. The free energy landscape is strongly modified and exhibits several minima for different ordered phases. As a result, the equilibrium phase depends on the temperature and phase transitions between macrophases appear. Additionally, minima that do not correspond to the canonical equilibrium give rise to metastable states characterized by extremely long lifetimes.

\begin{figure}[!ht]
\centering
\begin{tabular}{c}
\epsfig{file=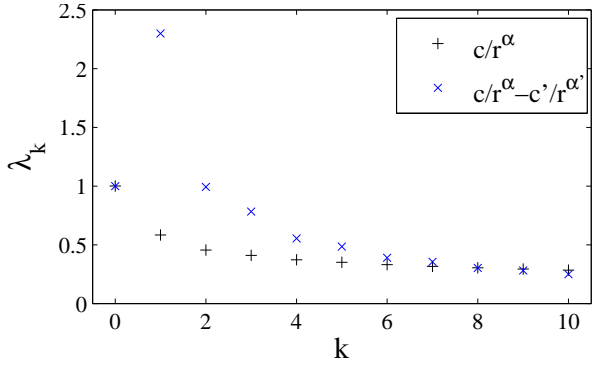,width=8cm}
\end{tabular}
\caption{Spectrum of a single power-law potential (black plus), with $\alpha=0.7$, and of the two competing power-law potentials under study (blue crosses), with $\alpha=0.1$ and $\alpha'=0$.\label{fig:Spectrum}}
\end{figure}

To investigate this force competition, an Ising-like model of $N$ spins on a lattice is used, inspired from the Dyson model with a single force~\cite{Dyson69a,Dyson69b,Thouless69}. It is a simple long-range model that captures the main features of force competition, and it is represented by the Hamiltonian
\begin{equation}
H=-\sum_{i,j} J_{ij}\sigma_i\sigma_j.\label{eq:Hij}
\end{equation}
Each spin can take values $\sigma_i=\pm1$ and is at a position $r_i=i/N$ on a lattice. The coupling between two spins, $J_{ij}$ is a function of their mutual distance, and can be characterized by a competition of power laws
\begin{equation}
J_{ij}=V(d_{ij})=\frac{c_{\alpha_1}}{\|i-j\|^{\alpha_1}}-\frac{c_{\alpha_2}}{\|i-j\|^{\alpha_2}}.
\end{equation}
Periodic boundary conditions are considered $\|i-j\|=\min(|r_i-r_j|,1-|r_i-r_j|)$, though the general conclusions of this work should remain valid for other boundary conditions. Using the eigenbasis on the potential (i.e. the Fourier basis)~\cite{Edlund10} the Hamiltonian (\ref{eq:Hij}) reads
\begin{equation}
H=-\frac{N}{2}\sum_k \lambda_k (\widehat\sigma_{c,k}^2+\widehat\sigma_{s,k}^2),\label{eq:Hk}
\end{equation}
where the eigenmodes and eigenvalues are
\begin{equation} \begin{array}{rcl}
\widehat\sigma_{c,k} &=& N^{-1}\sum_{i}\cos{(2\pi k r_i)} \sigma_i,
\\ \widehat\sigma_{s,k} &=& N^{-1}\sum_{i}\sin{(2\pi k r_i)} \sigma_i,
\\ \lambda_k&=&N^{-1}\sum_i \cos{(2\pi k r_i)} J_{1,i}.
\end{array} \end{equation}
A single power-law potential yields a monotically decreasing spectrum, but two competing ones can break this monotonicity, and allow for the emergence of modulated phases~\cite{Edlund10} shown in Fig.\ref{fig:Spectrum}.
The partition function, $Z=\sum_{\sigma_i=\pm1}\exp(-\beta H)$, contains a quadradic dependence on $\widehat\sigma$ in the exponential, which can be turned into a linear one using the Hubbard-Stratanovitch trick
\begin{equation}
\exp\left(\frac{N\beta\lambda \widehat\sigma^2}{2}\right)=\sqrt{\frac{N\beta\lambda}{2\pi}}\int_{-\infty}^\infty dy \exp\left(-\frac{N\beta\lambda y^2}{2}+N\beta\lambda y\widehat\sigma\right),\label{eq:HS}
\end{equation}
where $\beta=T^{-1}$. Considering the $K$ first Fourier modes, with $K$ a cut-off on the Fourier modes at an arbitrary order, and applying Eq.(\ref{eq:HS}) to them
\begin{widetext}
\begin{equation}
Z = \frac{1}{\Lambda}\left(\frac{N\beta}{2\pi}\right)^{K/2}\sum_{\sigma_i=\pm1}\int\mbox{d} y_k\mbox{d} y'_k \exp\Bigg[-\frac{ N\beta}{2}\sum_{k}\lambda_k(y_k^2+y_k^{\prime 2})
+ \beta\sum_i \sigma_i\sum_k\lambda_k\left(y_k \cos{(2\pi k r_i)}+y'_k \sin{(2\pi k r_i)}\right)\Bigg],\label{eq:Z1s}
\end{equation}
\end{widetext}
with $\Lambda=\prod_{k=0}^K\lambda_k$. Polar coordinates, $(y_k=\rho_k\cos\varphi_k,y'_k=\rho_k\sin\varphi_k)$, are introduced and the sum over the configurations of spins is performed
\begin{eqnarray}
Z &=& \frac{1}{\Lambda}\left(\frac{N\beta}{2\pi}\right)^{K/2}\int \rho_k\mbox{d} \rho_k\mbox{d} \varphi_k \exp\Bigg[-\frac{ N\beta}{2}\sum_{k}\lambda_k y_k^2 \nonumber
\\ &&+\sum_i\log \cosh\left(\beta\sum_k \lambda_k y_k \cos{(2\pi k r_i-\varphi_k)}\right)\Bigg].
\end{eqnarray}
It is noteworthy that the quantity in square bracket is extensive, so $Z$ can be written as $Z\propto \int\mbox{d} V\exp[-N\tilde\phi(\beta,\rho_k,\varphi_k)]$ where $\tilde\phi$ a function that is well-defined in the thermodynamic limit
\begin{equation} \begin{array}{rcl}
&&\tilde\phi(\rho_k,\varphi_k)=\frac{\beta}{2}\sum_{k}\lambda_k \rho_k^2 \label{eq:tphi}
\\ &&-\int_{-1/2}^{1/2}\log \cosh\big(\beta \sum_k \lambda_k\rho_k \cos{(2\pi k r-\varphi_k)}\big)\mbox{d} r.\nonumber
\end{array} \end{equation}
A similar treatment can be performed for short-range interactions using a mean-field approximation, but in the long-range case, it is an exact result. Using to the saddle-point method, in the thermodynamic limit, the canonical equilibrium is given by the minima of the function $\tilde\phi$
\begin{equation}
\rho_k=\int_{-1/2}^{1/2}\mbox{d} r\cos{(2\pi k r)}
\tanh\left(\beta\sum_{k'} \lambda_{k'}\rho_{k'} \cos{(2\pi k' r)}\right),\label{eq:sadrhok}
\end{equation}
for each mode $k$. We have here omitted the dependence on $\varphi_k$, since the equilibrium is degenerate on this variable. Indeed, for periodic boundary conditions, the particles are not distinguishable and the equilibrium has a translational degeneracy.

In general, Eq.(\ref{eq:sadrhok}) yields solutions that are a combination of eigenmodes. For the sake of simplicity, let us first consider a potential with only two modes $k=0,\ 1$. With the eigenvalues restricted by $\lambda_{k\geq2}=0$, the minima are determined by Eq.(\ref{eq:sadrhok}). For $k=0,\ 1$, the set of two equations is closed, therefore the amplitude of the $k\geq2$ modes may be derived from these. The function $\tilde\phi$ has two minima, $\overline\rho_{0}$ and $\overline\rho_{1}$, which are associated to each mode
\begin{subequations}\label{eq:rhob01}
\begin{eqnarray}
\overline\rho_0&=&\tanh\left(\beta\lambda_0\overline\rho_0\right),
\\ \overline\rho_1&=&\int_{-1/2}^{1/2}\mbox{d} r\cos{(2\pi r)}\tanh\left(\beta\lambda_1\overline\rho_1 \cos{(2\pi r)}\right).
\end{eqnarray}
\end{subequations}
In particular, using Eq.(\ref{eq:rhob01}), one can show that the equilibrium value of eigenmode $k$ is given by:
\begin{equation}
\langle \widehat\sigma_{k}^2\rangle=\frac{2}{\beta N}\frac{\partial \log Z}{\partial \lambda_k}=\overline\rho_k^2.
\end{equation}
Here, $\overline\rho_k$ must correspond to the minimum of the canonical equilibrium.

In the case of a single power-law potential, $V(r)\sim 1/r^\alpha$, the series of the $\lambda_k$ monotically decreases and the absolute minimum is associated to the mean-field mode, differing from above the critical temperature where the system remains in a disordered phase (i.e. $\rho_k=0,\ \forall k$)~\cite{Campa00}. However, when power-law forces are competing, provided $\lambda_1>2\lambda_0$, the absolute minimum yielded by (\ref{eq:rhob01}) depends on the temperature. In fact, the critical temperature below which a mode can exist is given by $T_k^c=\lambda_k(1+\delta_{0,k})/2$. Hence, if $\lambda_1>2\lambda_0$, the first Fourier mode exists at temperatures $T_0^c<T<T_1^c$ where the mean-field mode does not.

\begin{figure}[!ht]
\centering
\begin{tabular}{c}
\epsfig{file=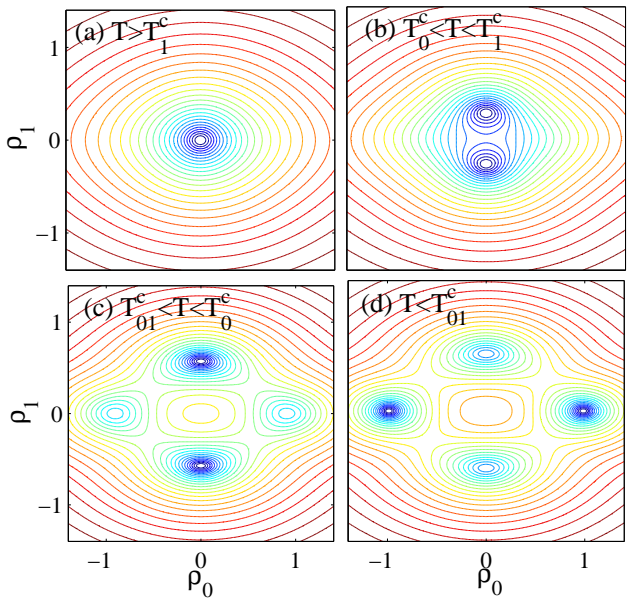,width=8.3cm}
\end{tabular}
\caption{Contour plot of $\tilde\phi$ (in log scale), in the $(\rho_0,\rho_1)$ space for (a) $T=1.4$, (b) $T=1.06$, (c) $T=0.6$ and (d) $T=0.4$. Simulations realized for $\lambda_0=1$ and $\lambda_1=2.3$, which yield  $T_1^c=1.15$, $T_0^c=1$ and $T_{01}^c\approx 0.459$. Although $\rho$ has been defined positive, it is also plotted for negative value, for the sake of visibility.\label{fig:profFT}}
\end{figure}

This phenomenon is illustrated in Fig.\ref{fig:profFT}, where the profile of $\tilde\phi$ is plotted. Above the critical temperature $T_1^c$, only the minimum corresponding to the disordered phase $\overline{\rho}_0=\overline{\rho}_1=0$ exists, Fig.\ref{fig:profFT}(a). Below $T_1^c$, but above $T_0^c$, a minimum associated to the first Fourier mode $\overline{\rho}_1>0$ appears and represents the statistical equilibrium, Fig.\ref{fig:profFT}(b). When the temperature reaches $T_0^c$, a minimum associated to the mean-field phase, $\overline{\rho}_0>0$ and $\overline{\rho}_1=0$, appears, but is still higher than the one associated with $k=1$ Fourier mode, while the latter remains the canonical equilibrium, Fig.\ref{fig:profFT}(c). However, at some temperature, $T_{01}^c$, the former minimum can turn into an absolute one and below this temperature, the mean-field mode remains the statistical equilibrium, Fig.\ref{fig:profFT}(d).

\begin{figure}[!ht]
\centering
\begin{tabular}{c}
\epsfig{file=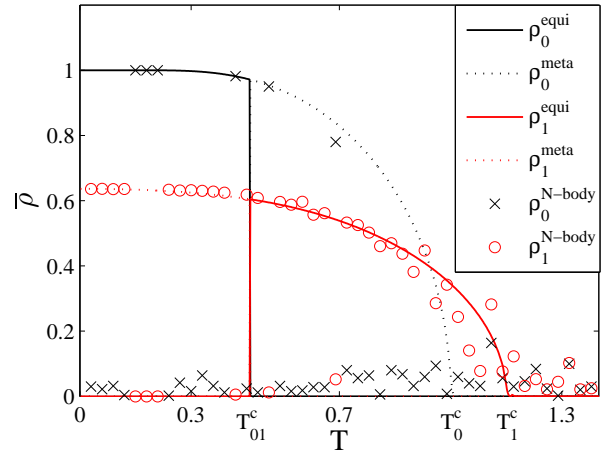,width=8cm}
\end{tabular}
\caption{Equilibrium magnetization for the $k=0$ (blue) and $k=1$ (red) modes for the truncated potential: the plain lines stand for the equilibrium curves, the dotted ones for the metastable states. The symbols correspond to microscopic simulations for the {\it complete} potential, with $N=1000$ and $N_{MC}=10^5$ Monte-Carlo steps.\label{fig:Equi}}
\end{figure}

Thus if the apparition of the minimum for the first Fourier mode corresponds to a second-order transition at $T=T_1^c$, then at $T=T_0^c$ the minimum associated with the mean-field mode appears in addition to a saddle point between the two minima (spinodal point). Eventually, the mean-field minimum turns into an absolute one at $T=T_{01}^c$, and the canonical equilibrium correspond to a new branch. At $T=T_{01}^c$, the slope of $\tilde\phi$ changes during the exchange of branch, which is a signature of a first-order transition. From a macroscopic point of view, the system jumps from a modulated phase to a homogeneous magnetized one, as can be seen from the plain lines in Fig.\ref{fig:Equi}. In this figure, the local minima of $\tilde\phi$ are metastable states (dotted lines). A mean-field metastable state appears when  $T$ goes below $T_0^c$ and the spatially-modulated mode turns metastable as the mean-field mode becomes the canonical equilibrium, at $T_{01}^c$.
 
Molecular simulations confirm these results, as is shown by the symbols in Fig.\ref{fig:Equi}. Simulations up to $10^8$ Monte-Carlo steps showed that the system either falls into the mean-field or the $k=1$ phase and is unable to escape the potential well it initially falls in. The simulations were realized with the {\it complete} potential, without truncation, which confirms that the $2$-mode truncation approach accepts the main feature of the new configuration. Modulated phases may also appear in the case of long-range magnetization-conserved dynamics~\cite{Mori11}.
 
The metastable states are considered in more detail here. The time needed for a long-range system to escape the metastable state is known to be a function of the exponential of $N$ times the barrier in free energy $\delta\tilde\phi$ that the system has to overcome~\cite{Griffiths66}. This barrier here corresponds to the difference between the saddle and the metastable states. Entropic barriers of order $10^{-4}-10^{-3}$ were reported in systems of coupled rotators~\cite{Antoni04}, while here barriers are hundreds times higher, $\delta\tilde\phi\approx 0.3$ for $T=0.3$, making the metastable extremely long-lived. Therefore the dynamics of systems with a thousand of particles remains trapped in the minimum of free energy they initially fall in for very large times, beyond what can be achieved today with a desktop computer. Fig.\ref{fig:Equi} illustrates the fact that molecular simuations predict the correct amplitude of the equilibrium and metastable states (blue and red crosses) up to finite-$N$ fluctuations, yet each realization stayed in the macroscopic state it initially falls in, even for long time simulations ($10^8$ Monte-Carlo steps).

Furthermore, for low temperatures $T<0.5$, the two minima are very close compared to the barrier that separate them, so that the typical time for the system to hop from the canonical equilibrium to the metastable state is of the same order as the time for the reverse process. Thus, the canonical equilibrium may be very difficult to identify using the time series of order parameters such as the population of the mean-field or first Fourier modes.

The present derivation holds for this Ising-like model, yet it can easily be generalized to models with continuous variables such as systems of rotators with power-law potentials~\cite{Anteneodo98}
\begin{equation}
H=\sum_i \frac{p_i^2}{2}-\frac{1}{2N}\sum_{i,j}V(d_{ij})\cos(\theta_i-\theta_j),
\end{equation}
where $(\theta_i,p_i)$ are conjugate variables. Following a similar derivation, one can show that the hyperbolic cosine in the partition function and in the free energy is replaced by the modified Bessel function, $I_0$~\cite{Campa00,Campa09}. Correspondingly in Eq.(\ref{eq:sadrhok}), the hyperbolic tangent should be replaced by $I_1/I_0$.
This set of equations yields the same phenomenology as for the modified Ising model, with the possible presence of modulated phases at the canonical equilibrium. The critical temperature of each mode is given by $T_k^c=\lambda_k(1+\delta_{0,k})/4$, showing that systems with continuous variables may as well exhibit modulated phases in the presence of competing potentials.

Finally, considering a more complex case, where $2\lambda_0<\lambda_{1}<\lambda_{2k+1}$ for $k'>1$, three different ordered phases may exist at equilibrium and are associated with the mean-field phase and to modes $1$ and $k'$~\footnote{Note that since the tanh is an odd function, Eq.(\ref{eq:sadrhok}) leaves the odd and even modes uncoupled.}. As before, the system switches from one ordered phase to the other via a first-order transition.

Note that the well of free energy the system preferentially falls in initially, starting from a disordered phase, can be deduced from an analysis of linear stability~\cite{Bachelard11} where each Fourier mode was shown to grow, in the linear regime, at an exponential rate that increases with $\lambda_k$. This growth rate, combined to finite-$N$ fluctuations, decides in which ordered phase the system initially falls. This explains why in Fig.\ref{fig:Equi} more simulations go to the modulated phase. Simulations with a lower ratio $\lambda_1/\lambda_0$ confirmed that the system distributes more equally between each phase during many realizations.

In conclusion, we have shown that the competition of long-range forces leads to new ordered phases, modulated at the scale of the system. The canonical equilibrium is characterized by first-order phase transition where the system switches from one phase to the other, yet the different ordered phases also present a strong metastable character over such long times that they may be confused with the canonical equilibrium.

The authors acknowledge fruitful discussions with J. Barr\'e, G. de Ninno, and especially S. Ruffo. R.B. acknowledge support from the Funda\c{c}\~ao de Amparo à Pesquisa do Estado de S\~ao Paulo (FAPESP).

\end{document}